\documentclass[aps,prd,twocolumn,nofootinbib]{revtex4}
\usepackage{graphicx}
\usepackage{color}
\usepackage{epstopdf}

\def\gr{$\gamma$-ray}

\begin{document}
\title{Sensitivity of $\gamma$-ray telescopes for detection of magnetic fields in intergalactic medium}
\author{A.~Neronov}
\affiliation{ISDC Data Center for Astrophysics, Chemin d'Ecogia 16, 1290 Versoix,
Switzerland and Geneva Observatory, 51 ch. des Maillettes, CH-1290 Sauverny,
Switzerland}
\author{D.V.~Semikoz}
\affiliation{APC, 10 rue Alice Domon et Leonie Duquet, F-75205 Paris Cedex 13, France}
\affiliation{Institute for Nuclear Research RAS, 60th October Anniversary prosp. 7a,
Moscow, 117312, Russia}

\begin{abstract}
We explore potential of current and next-generation \gr\ telescopes for the detection of weak magnetic fields in the intergalactic medium. We demonstrate that using two complementary techniques, observation of extended emission around point sources and observation of time delays in \gr\ flares, one would be able to probe most of the cosmologically and astrophysically interesting part of the  "magnetic field strength" vs. "correlation length" parameter space. This implies that  \gr\ observations with {\it Fermi} and ground-based Cherenkov telescopes will allow to (a) strongly constrain theories of the origin of magnetic fields in galaxies and galaxy clusters and (b) discover, constrain or rule out the existence of weak primordial magnetic field generated at different stages of evolution of the Early Universe.    
\end{abstract}
\maketitle

\section{Introduction}

Magnetic fields are known to play an important role in the physics of a variety of astrophysical objects, from stars to galaxies and galaxy clusters. The existence of galactic magnetic fields with strengths in the range $1-10\ \mu$G is established via observations of Faraday rotation and Zeeman splitting of atomic lines in the radio band and of polarization of starlight in the optical band \cite{kulsrud,beck08}. Magnetic fields of similar strength are found in the cores of galaxy clusters \cite{carilli02}. Weaker magnetic fields with strengths in the range $10^{-8}-10^{-7}$~G were recently discovered at the outskirts of galaxy clusters \cite{xu06,kronberg07}.

Although the strength and spatial structure of magnetic fields in the Milky Way and some other galaxies are reasonably well known today, there is no commonly accepted theory about the origin of these magnetic fields (see \cite{kronberg94,grasso00,widrow02,kulsrud} for recent reviews of the subject). There is a general agreement that the observed microGauss magnetic fields are the result of amplification of weak "seed" fields. The amplification mechanisms under discussion are the so-called "$\alpha-\omega$" dynamo (in the case of spiral galaxies) and/or compression and turbulent motions of plasma during the galaxy/cluster formation processes.

The nature of the initial weak seed fields for the dynamo or turbulent amplification is largely unknown \cite{kronberg94,grasso00,widrow02,kulsrud}. It might be that the seed fields are produced during the epoch of galaxy formation by electrical currents generated by the plasma experiencing gravitational collapse within a proto-galaxy \cite{pudritz89,gnedin00}, or ejected by the first supernovae \cite{rees87} or active galactic nuclei \cite{donnert09}. Otherwise, the seed fields might originate from still earlier epochs of the Universe expansion, down to the cosmological phase transitions or inflation times \cite{grasso00}.

Wide uncertainties in both the mechanism of amplification of the seed fields and in the nature of the seed fields themselves have led, over the last half-a-century, to the appearance of a long-standing problem of the "origin of cosmic magnetic fields" (in galaxies and galaxy clusters). 

It is clear that the clue for the solution of this problem might be given by the measurements of the initial seed fields. However, up to recently there was little hope that the extremely weak fields  outside galaxies and galaxy clusters would ever be detected. 

In what follows we show that direct measurements of the seed fields and derivation of constraints on their nature become possible with the newly available observations in the very-high-energy \gr\ band with space and ground-based \gr\ telescopes such as {\it Fermi}, HESS, MAGIC, VERITAS and, in the near future, CTA, AGIS and HAWC. The method of measurement of "ExtraGalactic" Magnetic Fields (EGMF) with \gr\ telescopes is based on the possibility of detection of emission from electromagnetic cascade initiated by the primary \gr s emitted by an extragalactic source and developing throughout the InterGalactic Medium (IGM) along the line of sight toward the source \cite{plaga,neronov07,japanese,elyiv09,kachelriess09}. Based on the knowledge of sensitivity of existing and future \gr\ telescopes, we find the range of EGMF parameters, such as the field strength $B$ and the correlation length $\lambda_B$, in which the EGMF is accessible for the measurements with one of the two available measurement techniques (imaging \cite{neronov07,elyiv09,kachelriess09} or timing \cite{plaga,japanese} of the cascade signal). We demonstrate that most of the astrophysically and cosmologically interesting range of EGMF parameters could be probed with \gr\ observations.

The plan of the paper is as follows. In Section \ref{sec:experiment} we summarize the existing bounds on the strength and correlation length of EGMF which come mostly from radio observations. In Section \ref{sec:cosmology} we discuss limits on the cosmological magnetic fields from cosmology. Then, in Section \ref{sec:theory} we compare the existing bounds to the theoretical predictions of two classes of models ("astrophysical" vs. "cosmological" models)  of the "seed" fields and show that model predictions normally fall largely below the existing bounds. In Sections \ref{sec:gammaray} -- \ref{sec:strongB} we first summarize the methods of measurement of EGMF with \gr\ telescopes and then estimate the ranges of EGMF parameters which can be probed with different observational techniques and different telescopes. Finally, in Section \ref{sec:conclusions} we draw conclusions from our study.

\section{Existing limits on the EGMF}
\label{sec:experiment}

\begin{figure}
\includegraphics[width=\linewidth]{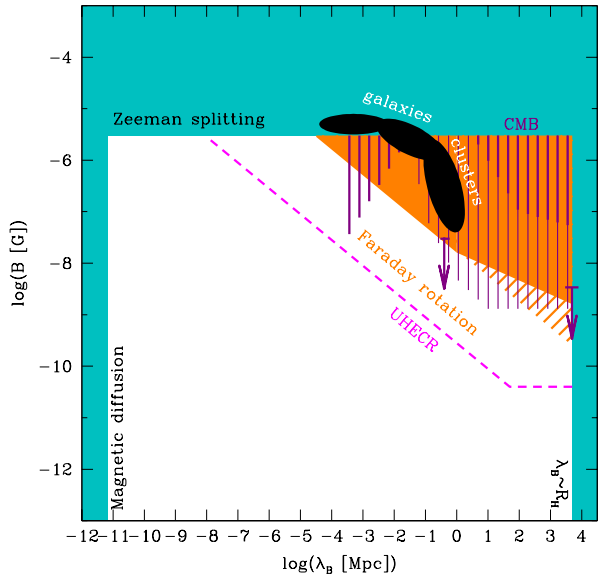}
\caption{Observational limits on EGMF. Cyan shaded region shows the upper limit on $B$ imposed by the Zeeman splitting measurement, the lower bound on the correlation length imposed by the magnetic diffusion and the upper bound on correlation length given by the Hubble radius. Orange shaded region shows the limit from Faraday rotation measurements. Filled orange region shows the limit derived in the Ref. \cite{blasi}, while the orange-hatched region is the limit derived in the Ref. \cite{kronberg76}.
Magenta line shows limit which can be imposed by observations of deflections of UHECR \cite{lee}. Violet vertical-hatched regions and the arrows at $\lambda_B\sim 0.5$~Mpc and $\lambda_B\sim R_H$ show the limits imposed on cosmologically produced fields by the CMB observations \cite{limitFIRAS,barrow,durrer00,Kahniashvili:2008hx}. Black ellipses show the ranges of measured magnetic fields in galaxies and galaxy clusters. }
\label{fig:exclusion_experiment}
\end{figure}

Contrary to the magnetic fields in galaxies and galaxy clusters, magnetic fields in the IGM have never been detected. Only upper limits, obtained by different observation techniques, exist. In this section we review the existing observational limits on the EGMF strength.

In the simplest settings, the EGMF configuration can be characterized by two parameters: the field strength, $B$, and the correlation length, $\lambda_B$\footnote{A third important parameter is the volume filling factor ${\cal V}$ of magnetic fields of a given strength $B$ and correlation length $\lambda_B$ Omitting this parameter we restrict ourself to the task of the search of the "dominant" EGMF, with volume filling factor ${\cal V}\sim 1$}. It turns out that limits on $B$ imposed by different observations depend on $\lambda_B$. This means that the limits could be presented as an "exclusion plot" in $(B,\lambda_B)$ parameter space, shown in Fig. \ref{fig:exclusion_experiment}.

Magnetic fields in IGM decay due to magnetic diffusion over the cosmological time on the distance scales \cite{grasso00}
\begin{equation}
\lambda_{\rm diff}=\sqrt{\frac{T_H}{4\pi\sigma}}\sim 2\times 10^{13}\mbox{ cm}
\end{equation}
where $T_H$ is the Hubble time and $\sigma\simeq 10^{11}$~s$^{-1}$ is the conductivity of the Universe after recombination. This means that the correlation length of EGMF is limited from below to $\lambda_B\ge \lambda_{\rm diff}$. At the same time, there are no known upper bounds on $\lambda_B$ and a natural bound is only set by the size of the visible part of the Universe, $\lambda_B\le R_H$, where $R_H$ is the Hubble radius. The lower and upper bounds on $\lambda_B$ are shown as vertical lines  in Fig. \ref{fig:exclusion_experiment}. 

\subsection{Zeeman splitting}

A straightforward upper bound on the EGMF strength can be found from the measurements of Zeeman splitting of 21 cm absorption line in the spectra of distant quasars, which are used to infer the magnetic field in the Milky Way galaxy \cite{zeeman}. The magnetic fields measured via Zeeman splitting are usually  in the range $1-100\ \mu$G and are commonly attributed to the field in the Milky Way \cite{zeeman} or in other galaxies (see e.g \cite{wolfe08} for detection of 84 $\mu$G magnetic field in a galaxy at redshift $z\simeq 0.7$). Measurements of $\sim \mu$G galactic magnetic fields via Zeeman splitting technique rule out the possibility of existence of still stronger magnetic fields in the IGM. The limit from Zeeman splitting measurements, obviously, does not depend on $\lambda_B$ and is shown as a horizontal (weakest) upper bound on $B$ in Fig. \ref{fig:exclusion_experiment}.

\subsection{Faraday rotation}

Measurements of Faraday rotation of polarized radio emission from distant quasars provide a possibility of detection of EGMF of the strength somewhat lower than the one accessible for the "direct" measurements via Zeeman splitting. The rotation measure $RM=\Delta\chi/\Delta\lambda^2$ ($\Delta \chi$ is the change of the polarization angle between the wavelengths $\lambda$ and $\lambda+\Delta\lambda$) is proportional to the product of magnetic field component along the line of sight, $B_{||}$ and the electron density of the IGM $n_e$ \cite{kronberg94}
\begin{equation}
\frac{\Delta \chi}{\Delta \lambda^2}\simeq 0.2^\circ {\cal F}(z_E)\left[\frac{B_{||,0}}{10^{-11}\mbox{ G}}\right]\left[\frac{n_{e,0}}{10^{-7}\mbox{ cm}^{-3}}\right]\mbox{ m}^{-2}
\end{equation}
where $z_E$ is redshift of the source of polarized emission (we use index $E$ for "emitter") and function ${\cal F}(z_E)=H_0\int_0^{z_E}(1+z)^3(dt/dz)dz$, if one assumes a simple redshift evolution of $B_{||}(z)\sim B_{||,0}(1+z)^2$, $n_e(z)=n_{e,0}(1+z)^3$ \cite{widrow02}.

In the $\Lambda$CDM cosmology, the derivative $dt/dz$ is given by
\begin{equation}
\label{dtau_dz}
\frac{dt}{dz} = \frac{1}{H_0(1+z)} \frac{1}{\sqrt{\Omega_M (1+z)^3+\Omega_\Lambda}}~~,
\label{dtdz}
\end{equation}
with $H_0\simeq 71$~km/s/Mpc,  $\Omega_M\simeq 0.27$ and  $\Omega_\Lambda\simeq 0.73$ being the present-day value of Hubble parameter, the cold dark matter and cosmological constant energy densities, respectively \cite{wmap}. 

Under simplest assumptions, free electrons are distributed homogeneously through the intergalactic medium. Using this model for electron distribution, a limit of $B\le 10^{-11}$~G  was obtained in the Ref. \cite{kronberg76} for the present-day electron density close to the critical density of the Universe. Re-scaling this limit for the electron density of the order of the baryon density, one would find a limit which is a right-bottom corner of the hatched orange region in Fig. \ref{fig:exclusion_experiment}.

The limit on EGMF imposed by the Faraday rotation measurements depends on the EGMF correlation length $\lambda_B$. If $\lambda_B\ll R_H$, the polarization angle experiences random changes due to the passages of multiple "cells" of the size $R\sim \lambda_B$ with coherent magnetic field. This means that the $\chi$ changes proportionally to the square root of the distance, 
\begin{equation}
\frac{\Delta \chi}{\Delta \lambda^2}\sim \sqrt{D\lambda_B}
\end{equation}
where $D$ is the distance to the radio source. Non-detection of the EGMF induced Faraday rotation implies, therefore, an upper limits on $B$ which scales as $B\sim \lambda_B^{-1/2}$. This upper limit is shown in Fig. \ref{fig:exclusion_experiment} as an orange-hatched region.

The simplest assumptions about homogeneous distribution of free electrons in the IGM might be an oversimplification. More complicated models of electron distribution were considered in the Refs. \cite{kronberg_perry} and \cite{blasi}. In these references IGM models based on the  Ly$\alpha$ data were considered. Under certain assumptions about the dependence of $n_e$ and $B$ on the density of the Ly$\alpha$ clouds, the authors of Ref. \cite{blasi} derive a limit on $B$ which is weaker than the one cited above (shown as a filled orange region in Fig. \ref{fig:exclusion_experiment}). 
The limit on $B_{||}$ depends weakly on $\lambda_B$ if $\lambda_B\gg \lambda_J$, where $\lambda_J$ is the Jeans length scale, which characterizes typical distance between the Ly$\alpha$ clouds \cite{blasi}. At small correlation lengths, $\lambda_B\ll \lambda_J$, the limit scales as $B_{||}\sim \lambda_B^{-1/2}$ since the Faraday rotation angle experience random changes during the passage through each cloud. 

\subsection{Deflections of UHECR}

Magnetic fields in IGM could be probed by the measurements of their effect on trajectories of charged particles (high-energy electrons and cosmic ray protons or nuclei), if their sources are known. Deflections of high-energy electrons and positrons are discussed below in Section \ref{sec:gammaray}.  In principle, measurements of deflections of  Ultra-High Energy Cosmic Rays (UHECR) with energies $E_{\rm UHECR}>10^{19}$ eV also can be used to constrain the intergalactic magnetic fields. In the simplest case, when the energy losses of UHECR could be neglected, the deflection angle of UHECR in a  regular magnetic field with coherence length $\lambda_B$ larger than the distance to the source $D$ is given by \cite{lee}:
\begin{eqnarray}
\label{UHECR_reg}
\theta_{\rm EGMF}&\simeq& \frac{ZeB_{\bot} D}{E_{\rm UHECR}}\\ &\simeq& 2.6^\circ Z\left[\frac{E_{\rm UHECR}}{10^{20}\mbox{ eV}}\right]^{-1}\left[\frac{B_\bot}{10^{-10}\mbox{ G}}\right]\left[\frac{D}{50\mbox{ Mpc}}\right]~,\nonumber
\end{eqnarray}
where where $B_\bot$ is the magnetic field component orthogonal to the line of sight, $E_{\rm UHECR}$ is particle energy and $Z$ is the atomic charge.

In an opposite case  $\lambda_B \ll  D$ one has \cite{lee}:

\begin{eqnarray}
\label{UHECR_turb}
\theta_{\rm EGMF}&\simeq&\frac{2}{\pi} \frac{ZeB_{\bot} \sqrt{D \lambda_B}}{E_{\rm UHECR}}\simeq 0.23^\circ Z\left[\frac{E_{\rm UHECR}}{10^{20}\mbox{ eV}}\right]^{-1}\\ && \left[\frac{B_\bot}{10^{-10}\mbox{ G}}\right]\left[\frac{D}{50\mbox{ Mpc}}\right]^{1/2}\left[\frac{\lambda_B}{1\mbox{ Mpc}}\right]^{1/2}
\nonumber
\end{eqnarray}
The main energy loss channels of UHECR protons is pion
production on CMB~\cite{GZK}, while for heavy nuclei it is photo-disintegration on cosmic infrared background \cite{stecker69}.
In both cases the energy / charge attenuation distances are $< 100$~Mpc. This limits the distances toward the sources of the highest energy cosmic rays to be not larger than $D\sim 100$~Mpc.

It is clear that measurement of EGMF with UHECR would be possible only under the condition that extragalactic point sources of UHECR would be detected (which is not the case at present). Even if the extragalactic UHECR point sources would be known, attempts of the measurement of EGMF would face the following principal difficulty.  For a known UHECR source, deflections of UHECR arrival directions from the real source position are determined not only by EGMF, but also by the deflections in the Galaxy, by magnetic fields in the intervening large scale structure elements, like galaxies or galaxy clusters along the line of sight and by deflections in the source host object (galaxy, galaxy cluster). As a result, the deflection angle $\theta$ is a sum of at least three terms,
\begin{equation}
\label{UHECR_total}
\theta =  \theta_{\rm Gal} + \theta_{\rm EGMF} +\theta_{\rm Source}~.
\end{equation}
Measurement of deflections by EGMF, $\theta_{\rm EGMF}$, via measurement of $\theta$ implies that the deviation by the Galactic magnetic field, $\theta_{\rm Gal}$ and possible deviations by the source host or intervening galaxies / galaxy cluster(s), $\theta_{\rm Source}$, are known. 

The magnetic field of the Milky Way galaxy is conventionally modeled as a sum of the regular and turbulent components of the field in the disk and halo of the Galaxy. This means that the term $\theta_{\rm Gal}$ in the above decomposition of $\theta$ is itself a superposition of at least four terms:
\begin{equation}
\label{UHECR_gal}
\theta_{Gal} =  \theta_{\rm Disk}^{\rm regular} + \theta_{\rm Disk}^{\rm turbulent} +\theta_{\rm Halo}^{\rm regular}+\theta_{\rm Halo}^{\rm turbulent}~.
\end{equation}

Deflection by regular and turbulent components of Galactic Disk and Halo, $\theta^{\rm Disk}_{\rm regular},\ \theta^{\rm Halo}_{\rm regular}$ and $\theta^{\rm Disk}_{\rm turbulent},\ \theta^{\rm Halo}_{\rm turbulent}$ , can be estimated by substitution of the typical disk/halo size at the place of $D$ and of the typical disk/halo field strength and correlation lengths at the place of $B, \lambda_B$ in Eqs. (\ref{UHECR_reg}), (\ref{UHECR_turb}).  Deflections of UHECR by the regular field in the disk $\theta_{\rm Disk}^{\rm regular}$ were studied in many theoretical models 
staring from Ref. \cite{Stanev:1996qj}. Typical values of parameters entering the analog of Eq.~(\ref{UHECR_reg}) imply substitution $D\rightarrow 2$ kpc length scale (for sources located far from the Galactic Plane) and  $B\rightarrow 2\ \mu$G, which give for   Eq.~(\ref{UHECR_reg}) $\theta_{\rm Disk}^{\rm regular} \simeq 4^\circ$. Turbulent fields are typically assumed to have coherence  scale $\sim 50$ pc and $B\simeq 4\ \mu$G, which, for  same field height scale $\sim 2$ kpc gives  $\theta_{\rm Disk}^{\rm turbulent} \simeq 0.5^\circ$.  
Contributions of the Halo fields are less certain, but result in deflections at least of the same order, see
recent discussion of all components in Refs. \cite{Sun:2007mx,Han:2009ts,Jansson:2009ip}. 

Although deflection angles of UHECR by the Galactic magnetic field can be readily estimated by order of magnitude, uncertainties in the measurements of Galactic magnetic field and discrepancies between the existing measurements and existing theoretical models of Galactic fields \cite{Sun:2007mx,Jansson:2009ip} do not allow to predict the deflection angle and direction of deflection for particular lines of sight (toward UHECR sources).  This means that, most probably, the details of the structure of the Galactic magnetic field along the line of sight toward UHECR sources would have to be deduced from the UHECR data itself, rather than just taken into account in the UHECR data analysis~\cite{Giacinti:2009fy}. This, obviously, will introduce a large uncertainty into the derivation of the properties of EGMF from the UHECR data.

Finally, the last term in the Eq.~(\ref{UHECR_total}),   $\theta_{\rm Source}$, is equally uncertain. Recent attempts of modeling of deflections of UHECR by the intervening elements of large scale structure, such as  galaxy clusters and/or filaments by two groups (see Refs. ~\cite{Sigl:2004yk} and ~\cite{Dolag:2004kp}) give contradictory results, which reflects uncertainties of the structure of magnetic fields inside and around clusters and filaments.  In addition, if an UHECR source is nearby, the source host galaxy or galaxy cluster could span several degrees on the sky. Significant deflections of UHECR by magnetic fields in the host galaxy or galaxy cluster could produce intrinsic 1-10 degree-scale extension of the source \cite{Dolag:2008py}, which would make the extraction of information about EGMF from the study of deflections of UHECR arrival directions from the source position still more problematic. 
  
Neglecting the above-mentioned problems, one could derive a theoretical "sensitivity" limit of future UHECR experiments for the detection of EGMF. Taking into account the typical  angular resolution of current and next generation UHECR experiments, like JEM-EUSO is $\theta_{\rm PSF}\simeq 2^\circ$~\cite{EUSO}, one could find, from Eqs. (\ref{UHECR_reg}) and (\ref{UHECR_turb}) that EGMF with the strength down to $B\sim 10^{-10}$~G (in the case of large correlation length $\lambda_B$) could influence the observational appearance of the signal from the UHECR source.  Theoretically possible sensitivity limit of UHECR experiments for the measurements of EGMF is shown as magenta line in Fig. \ref{fig:exclusion_experiment}.

\section{Limits on the cosmological magnetic fields from cosmology}
\label{sec:cosmology}

Zeeman splitting or Faraday rotation methods allow to detect, or put upper bounds, on the weak "seed " magnetic fields in the intergalactic medium in the present day Universe. As it is mentioned in Introduction, such fields could have been generated either during the epoch of galaxy formation, or at the earlier stages of evolution of the Universe. If the seed fields were generated before the epoch of recombination, additional constraints on the field strength and correlation length can be derived from the analysis of the Cosmic Microwave Background (CMB) data.

Most of the cosmological "magnetogenesis" models result in predictions of tangled magnetic field configurations with broad band power-law like spectra in Fourier space, cut off at a characteristic (model dependent) length scale $\lambda_B$, so that the Fourier components of the field have the form ~\cite{limitFIRAS}
\begin{equation}
\label{power_spectrum}
|B_k|^2 = B_0^2 \left(\frac{k}{k_B} \right)^n \frac{n+3}{4\pi} \Theta(k_B -k)~,
\end{equation}
where $k_B=2\pi/\lambda_B$ and $n$ is the power-law index. The energy density contained in the EGMF in the present day Universe,  $\rho_B^0=B_0^2/(8\pi)$, is expressed as an integral over the Fourier space $\rho_B^0= 1/(8\pi k_B^3) \int d^3k |B_k|^2$.
Note that, in order to have finite energy in magnetic field, one has to assume $n>-3$
in the Eq.~(\ref{power_spectrum}).

Constraints on the parameters of the magnetic field spectrum come from  the measurements of the anisotropies of the CMB spectrum, Faraday rotation of CMB, chemical potential of CMB and from Compton parameter measurements (analog of Sunyaev-Zel'dovich effect at recombination epoch). If the magnetic fields were generated prior to the epoch of Big Bang Nucleosynthesis (BBN), additional (weaker) constraints could be derived from the  BBN calculations, at the level of $B_0<10^{-6}$ G (see, for example, \cite{grasso00}). It is important to note that constraints on cosmological magnetic field strength $B$ normally depend not only on the correlation length  $\lambda_B$, but also on  unknown power law index $n$. 

The most straightforward limit on the strength magnetic fields homogeneous over the Hubble distance scale could be derived from the non-observation of the large angular scale anisotropies of the CMB. This gives an upper limit $B\le 4\times 10^{-9}$~G for the fields with $\lambda_B\sim R_H$ \cite{barrow}.  The powerlaw index dependent limit on the magnetic field strength were derived from the analysis of the CMB angular power spectrum in the Refs. \cite{durrer00,Giovannini:2008df, Giovannini:2009ts}. The envelope of the upper bounds on $(B,\lambda_B)$ for the range of the powerlaw indices $-3\le n\le 2$ is the lower boundary of the thin vertical violet hatched region of  $(B,\lambda_B)$ parameter space marked "CMB" in Fig. \ref{fig:exclusion_experiment}. 

Non-thermal dissipation of magnetic field energy into the energy of electrons/positrons during the recombination epoch could lead to distortion of blackbody CMB spectrum~\cite{CMBdamping}.   This distortion produces non-zero chemical potential $\mu$, which was calculated in the Ref.~\cite{limitFIRAS} in the form of a double integral. In order to calculate the bounds on $B,\lambda_B$, we consider two  limiting cases in which the integral could be taken analytically, namely, the cases when the magnetic field correlation length is much smaller or much larger than the characteristic damping length scale (the wavelength at which the magnetic field is damped by a factor $e$):
\begin{equation}
\label{damping}
\lambda_D = \frac{2 \pi}{z_\mu^{3/2}}\sqrt\frac{\overline{t}_0}{15 n_e^0 \sigma_{Th}} \approx 400\mbox{ pc},
\end{equation}
where time constant is $\overline{t}_0=2.4\times 10^{19}$ s, $z_\mu$ is the characteristic redshift of freeze-out from double-Compton scattering $z_\mu=2.5\times 10^{6}$,    $n_e^0$ is electron density and $ \sigma_{Th}$ is Thomson  cross section.

Taking into account the constraint  $|\mu|<9 \times 10^{-5}$ at 95 \% confidence level from COBE FIRAS data~\cite{FIRAS}, one can derive limits on magnetic fields in the analytical form in the two limiting cases. 
In the case  $\lambda_B \ll \lambda_D$ one has~\cite{limitFIRAS}:
\begin{equation}
\label{lim_low}
B < 3.2 \times 10^{-8} G \frac{1}{\sqrt{K}} \left(  \frac{\lambda_B}{400 ~\mbox{pc}}\right)^{-(n+3)/2}~,
\end{equation}
where  $K=1.4 \Gamma (n/2+5/2) \Gamma(3n/5 +9/5)2^{-(n+5)/2}(6/5)(n+3)$ is a constant of the order of unity, $K=0.8 / 2.1$ for $n=-2 / +1$. In the  opposite case $\lambda_B \gg \lambda_D$ the constraint is
\begin{equation}
\label{lim_high}
B < 3.2 \times 10^{-8} G \frac{1}{\sqrt{K_2}} \left(  \frac{\lambda_B}{400 ~\mbox{pc}}\right)~,
\end{equation}
where $K_2$ is another constant of the order of unity. The strongest limit on the field strength is obtained for the fields with correlation length $\lambda_B\sim\lambda_D$. For the fields with such correlation length no convenient analytical approximation could be found and instead a numerical integration of expression of the Ref.~\cite{limitFIRAS} should be performed. In Fig. \ref{fig:exclusion_experiment} we show the bound on $B,\lambda_B$ implied by the analysis of distortions of CMB spectrum as extrapolation of the analytical approximations given by Eq. (\ref{lim_low}), (\ref{lim_high}) for the whole ranges $\lambda_B<\lambda_D$, $\lambda_B>\lambda_D$. Note that the dependence of the limits on $B,\lambda_B$  on the power-law index $n$  practically disappears in the case  $\lambda_B\gg \lambda_D$ (only weak dependence remains in the constant $K_2$). At the same time, uncertainty in the value of $n$ "washes out" the upper bounds on $B$ at the length scales $\lambda_B\ll \lambda_D$.

We have cross-checked the above results by adopting an assumption that the magnetic field spectrum has the $\delta$-function rather than a power-law shape of Eq.(\ref{power_spectrum}).
In terms of Eq.~(\ref{power_spectrum}) this would correspond to the limit $n \rightarrow \infty$. As expected,
 Eq.~(\ref{lim_low})  does not give any constraints in this case for $\lambda_B<\lambda_D$, while
 limit Eq.(\ref{lim_high}) remains as it is.  
 
 Apart from producing a non-zero chemical potential, transfer of the magnetic field energy to electrons/ positrons could result in non-zero  Compton parameter $y$. Taking into account restrictions $y<1.5 \times 10^{-5}$ from COBE FIRAS, one finds a limit  $B<3  \times 10^{-8}$ G at $\lambda \sim 0.3-0.6 $ Mpc~\cite{limitFIRAS}. Note, that limits Eq.~(\ref{lim_low}) and  Eq.~(\ref{lim_high}) constrain magnetic fields created at $z>z_\mu=2.5\times 10^6$ ,
 while the limit following from restrictions on $y$ applies for fields created before recombination $z>2 \times 10^4$.

Magnetic fields created before recombination could produce another observable phenomenon: Faraday rotation of linear polarization of CMBR. Constraints on $B,\lambda_B$ stemming from non-observation of this effect were first discussed in the Ref.~\cite{KosowskyLoeb96} and subsequently updated  using the 5-years data of WMAP in the Ref.~\cite{WMAP_5years}. The limits coming from non-observation of Faraday rotation in CMB signal are, at present, weaker than the limits imposed by the rotation measures of distant blazars or limits from  the CMB angular power spectrum \cite{Kahniashvili:2008hx}. The limit from non-observation of Faraday rotation in the CMB signal is shown by thick violet vertically hatched region.

\section{Theoretical predictions}
\label{sec:theory}
\begin{figure*}
\includegraphics[width=\linewidth]{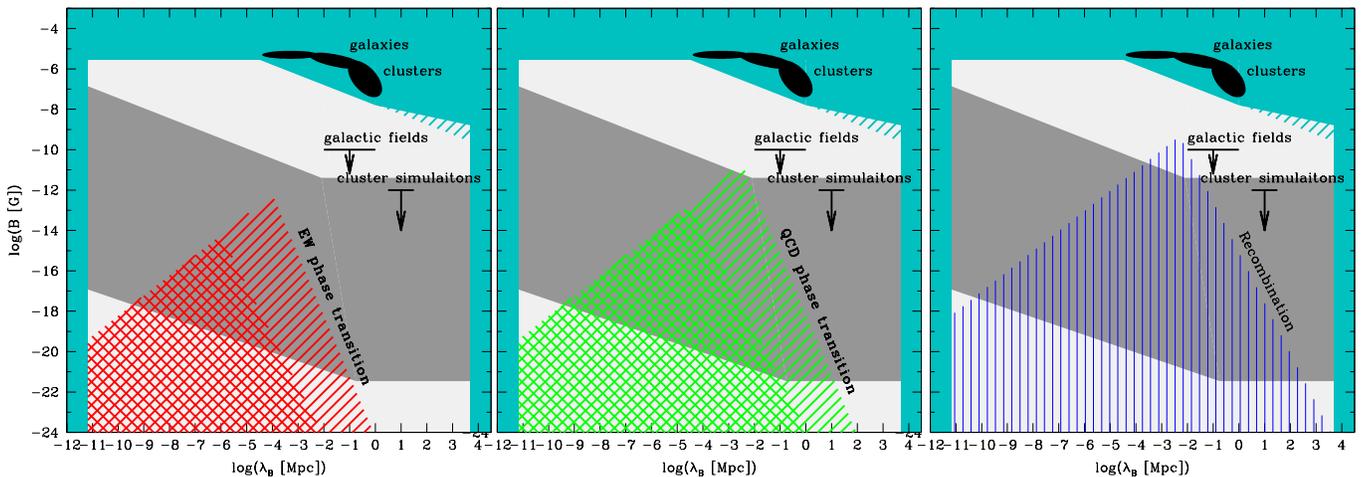}
\caption{Model predictions and estimates for the EGMF strength. Cyan shaded region and black ellipses show the experimental limits and measurements  from Fig. \ref{fig:exclusion_theory}. Upper bound at $B\sim 10^{-10}$~G shown by solid line comes from flux conservation during galaxy formation argument \cite{grasso00}. Upper bound at $B=10^{-12}$~G shows a limit imposed by constrained simulations of magnetic fields in galaxy clusters \cite{Dolag:2004kp,donnert09}. Left panel: left and right hatched regions show theoretically allowed range of values of $B,\lambda_B$ for non-helical and helical fields generated at the epoch of electroweak phase transition  during radiation-dominated era. MIddle panel: left and right hatched region show ranges of possible $B,\lambda_B$ for nonhelical and helical magnetic fields produced during the QCD phase transition. Right panel: hatched region is the range of possible $B,\lambda_B$  for EGMF generated during recombination epoch. Dark grey shaded region shows the range of $(B,\lambda_B)$ parameter space accessible for the \gr\ measurements via \gr\ observations. Light-grey shaded regions show the parts of the parameter space in which the existence of EGMF could be confirmed or ruled out, but no measurements of EGMF strength is possible.}
\label{fig:exclusion_theory}
\end{figure*}

Search for extremely weak EGMF is important in the context of the problem of the origin of magnetic fields in galaxies and galaxy clusters. Most commonly accepted hypothesis about the origin of magnetic fields in spiral galaxies is that they were produced via the so-called "$\alpha-\omega$ dynamo" mechanism from tiny "seed" fields of uncertain origin \cite{kulsrud}. Estimates of the efficiency of $\alpha-\omega$ dynamos imply that the "seed" field should have strength of the order of $\sim 10^{-21}$~G or higher, although these estimates suffer from large uncertainties\footnote{Recent discoveries of strong magnetic fields in galaxies at significant redshifts imply still stronger "seed" magnetic fields, see e.g. \cite{wolfe08}}. Alternatively, magnetic fields in galaxies and galaxy clusters could be produced via compression of the seed magnetic  fields present during structure formation epoch, and their amplification by the turbulence (see e.g. Ref. \cite{grasso00}).  Decisive observational test of alternative theories of the origin of cosmic magnetic fields can be given only by observations of the initial weak seed fields implied in all the theoretical models.

However, the nature of the weak seed fields  remains largely unconstrained. Theoretically, two main hypotheses exist: that of the "astrophysical origin"  (see e.g. \cite{kulsrud}) and of the "cosmological origin" (see e.g. \cite{grasso00}) of the seed fields. The "astrophysical origin" models usually assume that the seed fields are produced via the so-called "Biermann battery" effect (difference in mobility of electrons and protons in astrophysical plasma) at the early stages of galaxy formation. Different types of collective motions of plasma could be involved, such as the outflows from the first supernovae \cite{rees87}, activity of active galactic nuclei \cite{donnert09}, gravitational collapse \cite{pudritz89,gnedin00} and/or turbulence \cite{kraichnan67,ryu08} in proto-galaxies.   The "cosmological" models usually exploit a similar effect at earlier stages of evolution of the universe, during or before recombination \cite{harrison70,rebhan92}, electroweak \cite{baym94,vachaspati91,grasso98} and QCD \cite{quashnock89,cheng94,sigl97} phase transitions or at still earlier epochs \cite{turner88,dolgov93,ratra92,joyce97} (see \cite{grasso00} for a review). It is clear that detection of the seed magnetic fields would not only help to resolve the problem of the origin of magnetic fields in galaxies and galaxy clusters, but also provide a new observational data constraining physical conditions in the Early Universe.

Significant difference between the "cosmological origin" and "astrophysical origin" models of the seed fields is that in the former case  weak seed magnetic field should be present everywhere the Universe today. In particular, the seed fields could be found in the voids of large scale structure, outside galaxies and galaxy clusters. To the contrary, if the seed fields are produced via astrophysical mechanisms inside proto-galaxies, no magnetic field generation outside galaxies and clusters is expected. Magnetic field in the voids of the large scale structure is expected to be close to zero.

The "cosmological origin" models  could be divided onto two broad classes: models based on mechanisms operating during phase transitions in the Early Universe and models based on mechanisms operating during inflation epoch. In the latter case, the fields might be produced on a super-horizon scale and no firm theoretical limits on the characteristic field correlation length can be deduced. In the former case, the strength and correlation length of cosmologically generated magnetic fields are well limited from above. 

The correlation length of magnetic field can not exceed the size of the horizon at the moment of phase transition
\begin{equation}
\lambda_{B*}\le \frac{a_0}{a_*H_*}\sim\frac{1}{\sqrt{G_N}T_*T_0}\sim 10^{15}\left[\frac{T_*}{100\mbox{ GeV}}\right]\mbox{ cm}
\end{equation}
where $a,H$ and $T$ are  scale factor, expansion rate and temperature of the Universe and $G_N$ is the Newton constant. Indices $0$ and $*$ refer to the parameter values  today and at the moment of magnetic field production, respectively. 
The magnetic field strength at the correlation length $\lambda_B$ is limited by the requirement that the magnetic field energy density $\rho_B=B^2/(8\pi)$ should not overclose the Universe. Since the magnetic field energy density evolves in the same way as the radiation energy density, $\rho_B\sim \rho_{ph}\sim a^{-4}$, present day magnetic field at the scale $\lambda_B$ is limited to be below the field for which the energy density is equal to the CMB energy density, 
\begin{equation}
\label{bspectrum}
B(\lambda_{B*})\le \sqrt{4\pi g_*\rho_{\rm CMB}}\simeq 3\times 10^{-6}(g_*/2)^{1/2}\mbox{ G}
\label{B_CMB}
\end{equation}
where $g_*$ is the number of relativistic degrees of freedom at the moment of magnetogenesis. 
If the magnetic field would grow so that it would start to contribute significantly to the overall energy density of the Universe, its further amplification during the magnetogenesis epoch would stop because the energy transferred to the magnetic field would be dissipated via strong gravitational wave production \cite{durrer00,caprini}. 

The upper limit on the magnetic field strength at the horizon scale (\ref{bspectrum})   implies an upper limit on super-horizon scales which follow from straightforward causality arguments \cite{durrer}. 
For non-helical magnetic fields an upper limit on the strength of magnetic field at arbitrary length scale $\lambda$ could be obtained taking into account the fact that the power-law index of magnetic fields produced in a causal way at  phase transition(s) are limited to be $n\ge 2$ \cite{durrer} 
\begin{equation}
\label{bi1}
B(\lambda_B)\le B(\lambda_{B*})\left(\frac{\lambda_B}{\lambda_{B*}}\right)^{-(n+3)/2}\le 3\times 10^{-6}\left(\frac{\lambda_B}{\lambda_{B*}}\right)^{-5/2}
\end{equation}

Energy contained in magnetic fields with small enough correlation length is dissipated in the course of evolution of the universe. This leads to the increase of the "integral scale" (i.e. the distance scale which gives dominant contribution to the  magnetic magnetic field energy density) with time \cite{jedamzik04}
up to the scale 
\begin{equation}
\label{bi2}
\lambda_{B,I}\sim \frac{v}{H}
\end{equation}
where $v$ is the characteristic velocity scale, which is of the order of either Alfven or viscous velocity at different epochs of the Universe expansion. Numerically, $\lambda_{B,I}\sim 1\left[B/5\times 10^{-12}\mbox{ G}\right]\mbox{ kpc}$ for magnetic fields produced much before recombination epoch and $\lambda_{B,I}\sim 1\left[B/8\times 10^{-11}\mbox{ G}\right]\mbox{ kpc}$ for the magnetic fields produced at recombination. 

The above constraints limit the possible present day values of $B, \lambda_B$ for the magnetic fields produced in the Early Universe. The possible ranges of the $(B,\lambda_B)$ parameter space are shown in Fig. \ref{fig:exclusion_theory} for the cases when magnetogenesis proceeds during electroweak or QCD phase transitions or at the moment of recombination. These regions are bound on the left by the constraint on $\lambda_{B,I}$ given by Eqs. (\ref{bi1}), (\ref{bi2}). The right side boundaries of the allowed regions are determined by the constraints (\ref{bspectrum}), (\ref{bi1}). 

If the magnetic fields possess non-zero helicity, the energy contained in the short-wavelength modes can be transferred to the longer-wavelength modes via the development of "inverse cascade", rather than dissipated in the fluid motions. This results in a slower decay of the magnetic field energy density and faster growth of the integral scale, than in the case of non-helical fields \cite{jedamzik04}. 

For the helical magnetic fields, the magnetic field power can be transferred from the initial scale $\lambda_{B*}$ to a larger scales $\overline \lambda_{B}$ via inverse cascade. Conservation of helicity implies that ${\cal H}\sim B^2L$, where $L$ is the distance scale, is conserved during the inverse cascade. Analytical/numerical calculations of the inverse cascades show that the cascade lengthscale evolves as $aL\sim t^{2/3}$ \cite{jedamzik04} during the radiation dominated epoch, which means that 
\begin{eqnarray}
\label{bhelical}
\overline \lambda_{B}&\simeq& \left(\frac{T_*}{1\mbox{ eV}}\right)^{1/3}\lambda_{B*}\nonumber\\
\overline B(\overline\lambda_{B})&\simeq& \left(\frac{T_*}{1\mbox{ eV}}\right)^{-1/6}B(\lambda_{B*})
\end{eqnarray}
at the end of the radiation-dominated epoch at $T\sim 1$~eV. The causality argument applied for the helical magnetic fields gives the constraint on the power-law index $n\ge 3$ at the scales $\lambda\ge \overline\lambda_{B}$ \cite{durrer}. The limit (\ref{bhelical}) is weaker than the bound (\ref{bi1}) on non-helical magnetic fields. This limit determined the right-side boundary of the allowed regions of ($B,\lambda_B)$ parameter space for the electroweak and QCD phase transition magnetogenesis in Fig. \ref{fig:exclusion_theory}. 

It is interesting to note that predictions for the strength and correlation length of the "primordial" magnetic fields produced at electroweak and QCD phase transitions fall in a region of $(B,\lambda_B)$ parameter space which is not accessible for the existing measurement techniques, such as Faraday rotation or Zeeman splitting methods. However, it turns out that this region of $(B,\lambda_B)$ parameter space is accessible for the measurement techniques which exploit the potential of the newly opened field of very-high-energy (VHE) \gr\ astronomy. In the following sections we demonstrate that using the current and next generation ground and space based \gr\ telescopes one could probe the part of $(B,\lambda_B)$ parameter space shown by the light and dark grey-shaded regions in Fig. \ref{fig:exclusion_theory} and in this way test the models of the origin of seed fields and the models of the origin of magnetic fields in galaxies and galaxy clusters.

\section{Measurements of EGMF with \gr\ telescopes}
\label{sec:gammaray}

\subsection{Basic formulae}

Multi-TeV $\gamma$-rays emitted by distant point sources are not able to propagate over large
 distances because of the absorption in interactions with optical/infrared extragalactic background light (EBL).
 
Redshift-dependent inverse mean free path of such gamma-rays of the energy $E_{\gamma_0}'$  propagating at redshift $z$ through the EBL of the density $n_{EBL}(\epsilon,z)$ can be defined as 
\begin{eqnarray}
&& D_\gamma(E_{\gamma_0}',z)^{-1}= <\sigma_{\gamma\gamma} n_{\rm EBL}>  \nonumber \\  
&=&  \int_{\epsilon_{min}'}^\infty d\epsilon' \frac{dn_{EBL}(\epsilon' , z)}{d\epsilon'}  \int_{-1}^{1} d\mu  (1-\mu)\sigma_{\gamma\gamma}(s), 
\label{Dgamma00}
\end{eqnarray}
where $\sigma_{\gamma\gamma}(s)$ is pair production cross section which depends on  $s=2 E_{\gamma_0}' \epsilon' (1-\mu)$ with $\mu$ being cosine of the angle between the directions of propagation of \gr\ and background photon.  The limit of integration $\epsilon_{min}' = m_e^2/E_{\gamma_0}'$ in Eq.  (\ref{Dgamma00}) corresponds to the threshold of the pair production.

Estimates of $D_\gamma(E_{\gamma_0}',z)$ suffer from significant uncertainty in the modeling 
of $n_{EBL}(\epsilon , z)$. There exist several theoretical models for the formation of infrared/optical  background based on the models of evolution of starlight and dust emission from different types of galaxies at different redshifts. The resulting estimates of $n_{EBL}(\epsilon , z)$ widely differ both in the predicted
shape of local  $n_{EBL}(\epsilon , z=0)$ and in its evolution with increasing redshift, see  Refs.~\cite{kneiske04,franceschini08,Primack08,Stecker06}. However, all the models agree in a general trend of decrease of $D_\gamma$ with the increasing energy and redshift.  Following 
results of ref.~\cite{Raue:2008bw}, one can assume that $n_{EBL}(\epsilon , z) \approx  (1+z)^{-2} n_{EBL}(\epsilon, z=0)$. In this case the integrals in the Eq.~(\ref{Dgamma00}) depend on $z$ only through  $E_{\gamma_0}'$. Simplifying the expression (\ref{Dgamma00}) in this way, we have found that in the energy band of interest for the following discussion, a broad range of model predictions/uncertainties for $D_\gamma$ could be described by the following simple approximation
\begin{equation}
D_\gamma(E_{\gamma_0}',z) = 40\frac{\kappa }{(1+z)^2} \left[\frac{E_{\gamma_0}'}{20\mbox{ TeV}}\right]^{-1}\mbox{ Mpc}~,
\label{Dgamma0}
\end{equation}
where a numerical factor $\kappa=\kappa(E_{\gamma_0},z)\sim 1$ accounts for the model uncertainties.  Comparing the results of calculations of the Refs. \cite{kneiske04,franceschini08,Primack08,Stecker06} one can find that  in the energy range $E_{\gamma_0}'\sim 0.1-10$~TeV and the redshift range  $z<1$ the uncertainty in $\kappa$ is as large as $0.3\le\kappa\le 3$.  

As the \gr\ propagates from the source toward the Earth, the optical depth with respect to the pair production grows as 
\begin{equation}
\frac{d\tau}{dt}=\frac{1}{D_\gamma(E_{\gamma_0}',z)}
\end{equation}
$\tau$ reaches 1 at the time $t_{\gamma\gamma}$, which can be implicitly found from equation
\begin{equation}
\label{t_gammagamma}
\int_{t_E}^{t_{\gamma\gamma}}\frac{dt}{D_\gamma(E_{\gamma_0}',z)}=
\int_{z_E}^{z_{\gamma\gamma}}\frac{dz}{D_\gamma(E_{\gamma_0}',z)}\frac{dt}{dz}=1
\end{equation}
where $t_E$ is the time of emission of photon from the source and $z_E$ and $z_{\gamma\gamma}$ are, the redshifts corresponding to the times $t_E$ and $t_{\gamma\gamma}$.

In principle, the mean free path of very-high-energy \gr s from distant sources could be large enough so that $z_{\gamma\gamma}$ could be significantly different from $z_E$. In this case the exact expression  (\ref{t_gammagamma}) should be used to determine $t_{\gamma\gamma}$. The "comoving" mean free path $d_\gamma$ of \gr s of initial energy $E_{\gamma_0}'(z_E)$ could then be estimated as
\begin{equation}
\label{dg}
d_{\gamma}\left[E_{\gamma_0}'(z_E)\right]=\int_{t_E}^{t_{\gamma\gamma}}\frac{dt'}{a(t')}
\end{equation}
where $a(t)=a_0/(1+z)$.

In the following calculations (which are mostly order-of magnitude estimates) we will adopt a simplifying assumption that $z_{\gamma\gamma}\simeq z_E$. In this case $t_{\gamma\gamma}-t_E$ could be explicitly found from Eq. (\ref{t_gammagamma}) and expression (\ref{dg}) reduces to 
\begin{equation}
\label{dg1}
d_{\gamma}\left[E_{\gamma_0}'(z_E)\right]\simeq \frac{D_{\gamma}(E_{\gamma_0}',z_E)}{a(t_E)}.
\end{equation}
One should remember, however, that assumption $z_{\gamma\gamma}\simeq z_E$ does not hold for relatively low energy (sub-TeV) \gr s which could produce cascade signal in {\it Fermi} (GeV) energy band. 

Using Eq. (\ref{dtau_dz}) one can find the optical depth with respect to the pair production for the \gr s of the energy $E_{\gamma_0}$ that reach observer on the Earth:
\begin{equation}
\tau(E_{\gamma_0},z_E)=\int_0^{z_E} dz \frac{dt}{dz} \frac{1}{D_\gamma\left((1+z)E_{\gamma_0},z\right)}
\label{tau_exact}
\end{equation}

Adopting the approximation (\ref{Dgamma0}) for $D_{\gamma_0}$
one finds the estimate for $\tau$ 
\begin{equation}
\tau(E_{\gamma_0},z_E)=\frac{2(\sqrt{\Omega_M(1+z)^3+\Omega_\Lambda}-1)}{3H_0\Omega_MD_\gamma(E_{\gamma_0},0)}
\label{tau_we}
\end{equation}
Uncertainties in the model predictions for $D_\gamma(E_{\gamma_0}',z)$ result in the discrepancies between the model predictions for $\tau$. Comparing the results of different calculations reported in the literature, one can find, for example, $\tau=1$ 
at $z=0.03$  for $E_\gamma=5-18$ TeV in the models of ref.~\cite{kneiske04},  $E_\gamma=9$  TeV in the model
of \cite{franceschini08}, $E_\gamma=7-8$  GeV in the models of ref.~\cite{Primack08} and $E_\gamma=2.7$ TeV for
fast evolution model of ref.~\cite{Stecker06}, the value from Eq.~(\ref{tau_we})  is   $E_\gamma=6$  TeV.
One can verify that the estimate of $\tau$ given by Eq. (\ref{tau_we}) is within factor 2 from results of \cite{franceschini08}
for $z<1$ and $0.2 ~\mbox{TeV}~ < E_\gamma <30 ~\mbox{TeV}$. Taking into account the fact that error introduced by approximation (\ref{tau_we}) is smaller than uncertainty of the models,  we will use approximations of Eqs.~(\ref{Dgamma0}) and Eq.~(\ref{tau_we}) instead of exact Eqs.~(\ref{Dgamma00}) and Eq.~(\ref{tau_exact}) in the following sections.

\begin{figure}
\includegraphics[width=\linewidth]{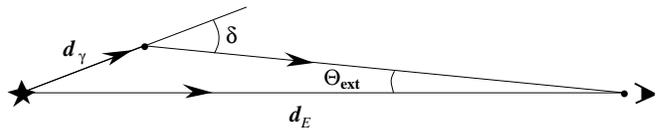}
\caption{Geometry of propagation of the direct and cascade \gr s from the source (on the left) to the observer (on the right).}
\label{fig:scheme}
\end{figure}
 
 The pair production on EBL reduces the flux of  $\gamma $-rays from the source by
\begin{equation}
\label{absorb}
F(E_{\gamma_0})=F_0(E'_{\gamma_0}(z_E))e^{-\tau(E_{\gamma_0},z_E)},
\end{equation}
where 
$F(E_{\gamma_0})$ is the detected spectrum, $F_0(E'_{\gamma_0})$ is the initial spectrum of the source and $\tau(E_{\gamma_0},z_E)$ is the optical depth (\ref{tau_exact}), (\ref{tau_we}). 

The $e^+e^-$ pairs 
 produced in interactions of multi-TeV $\gamma$-rays with EBL photons produce secondary $\gamma$-rays via inverse Compton (IC) scattering of the Cosmic Microwave Background (CMB) photons. Typical energies of the IC photons reaching the Earth are
\begin{equation}
E_{\gamma} = \frac{4}{3} (1+z_{\gamma\gamma})^{-1}\epsilon_{CMB}'\frac{E_e'^2}{m_e^2}\simeq 0.32
\left[\frac{E'_{\gamma_0}}{20 \mbox{ TeV}}\right]^2\mbox{ TeV}
\label{Esec}
\end{equation}
where $\epsilon_{CMB}'=6\times 10^{-4}(1+z_{\gamma\gamma})$~eV is the typical energy of CMB photons. In the above equation we have assumed that the energy of primary $\gamma$-ray is $E'_{\gamma_0}\simeq 2E_e'$ with $E_{\gamma_0}'$ being the energy of the primary \gr s at the redshift of the pair production. Upscattering of the infrared/optical background photons gives sub-dominant contribution to the IC scattering spectrum because the energy density of CMB is much higher than the density of the infrared/optical background.

 Deflections of $e^+e^-$ pairs produced by the $\gamma$-rays, which were initially emitted slightly
away from the observer, could lead to "redirection" of the secondary cascade photons toward the observer. This effect leads to the appearance of two potentially observable effects: extended emission around an initially point source of \gr s \cite{neronov07,elyiv09,kachelriess09} and delayed "echo" of \gr\ flares of extragalactic sources \cite{plaga,japanese}.

The cascade electrons loose their energy via IC scattering of the CMB photons within the distance
\begin{equation}
D_e=\frac{3m_e^2c^3}{4\sigma_TU_{\rm CMB}'E_e'}\simeq  10^{23}(1+z_{\gamma\gamma})^{-4}\left[\frac{E_e'}{10\mbox{ TeV}}\right]^{-1}\mbox{ cm}
\end{equation}
The deflection angle of the $e^+e^-$ pairs, accumulated over the cooling distance, depends on the correlation length of magnetic field, $\lambda_B$. 

If the correlation length $\lambda_B$ is much larger than $D_e$, motion of
 electrons and positrons at the length scale $D_e$ could be approximated by the motion in homogeneous magnetic field. In this case typical deflection angle $\delta$ is estimated as a ratio of $D_e$ to the  Larmor radius in magnetic field $B'$,
\begin{equation}
R_L=\frac{E_e'}{eB'}\simeq 3\times 10^{28}\left[\frac{B'}{10^{-18}\mbox{ G}}\right]^{-1}\left[\frac{E_e'}{10\mbox{ TeV}}\right]\mbox{ cm.}
\end{equation}
Note, that, in principle, EGMF depends on the redshift, $B'=B'(z)$. In the simplest case, when the magnetic field strength changes only in result of expansion of the Universe, $B'(z)\sim B_0(1+z)^2$, where $B_0$ is the present epoch EGMF strength.
This gives
\begin{eqnarray}
\delta&=&\frac{D_e}{R_L}\simeq 3\times 10^{-6}(1+z_{\gamma\gamma})^{-4}\left[\frac{B'}{10^{-18}\mbox{ G}}\right]\left[\frac{E_e'}{10\mbox{ TeV}}\right]^{-2}\nonumber\\
&\simeq& 3\times 10^{-6}(1+z_{\gamma\gamma})^{-2}\left[\frac{B_0}{10^{-18}\mbox{ G}}\right]\left[\frac{E_e'}{10\mbox{ TeV}}\right]^{-2}
\label{delta1}
\end{eqnarray}

If the correlation length of magnetic field, $\lambda_B'$, is much less than electron cooling distance $D_e$, electron deflections are described as diffusion in angle, so that the deflection angle is estimated as
\begin{eqnarray}
\delta&=&\frac{\sqrt{D_e\lambda_B}}{R_L}\simeq 5\times 10^{-7}(1+z_{\gamma\gamma})^{-2}\left[\frac{E_e'}{10\mbox{ TeV}}\right]^{-3/2}\nonumber\\ &&\left[\frac{B'}{10^{-18}\mbox{ G}}\right]\left[\frac{\lambda_B'}{1\mbox{ kpc}}\right]^{1/2}\\
&\simeq& 5\times 10^{-7}(1+z_{\gamma\gamma})^{-1/2}\left[\frac{E_e'}{10\mbox{ TeV}}\right]^{-3/2}\nonumber\\ &&\left[\frac{B_0}{10^{-18}\mbox{ G}}\right]\left[\frac{\lambda_{B0}}{1\mbox{ kpc}}\right]^{1/2}\nonumber
\label{delta2}
\end{eqnarray}
where we have assumed that $\lambda_B'$ scales with $z$ as $\lambda_B'=\lambda_{B0}(1+z)^{-1}$ with $\lambda_{B0}$ being the present epoch EGMF correlation length.

Knowing the deflection angle of electrons, one can readily find the angular extension of the secondary IC emission from the $e^+e^-$ pairs using simple geometrical calculation in the comoving reference system, shown in Fig. \ref{fig:scheme}. In this figure $d_E$ is the comoving distance to the source, defined as
\begin{equation}
d_E=\int_0^{t_E}\frac{dt}{a(t)}=\frac{1}{a_0H_0}\int_0^{z_E}\frac{1}{\sqrt{\Omega_M(1+z)^3+\Omega_\Lambda}}dz
\end{equation}
and $d_\gamma$ is the \gr\ mean free path given by Eq. (\ref{dg}) which, in the case $z_{\gamma\gamma}\simeq z_E$ can be calculated using Eq. (\ref{dg1}).
The angle $\Theta_{\rm ext}$ is expressed through $d_E, d_\gamma$ and $\delta$ as
\begin{equation}
\sin(\Theta_{\rm ext})=\frac{d_{\gamma}[E_{\gamma_0}']}{d_E}\sin\delta
\end{equation}

In the case $z=z_E\simeq z_{\gamma\gamma}$ the above expression reduces to
\begin{equation}
\Theta_{\rm ext} (E_\gamma)= \frac{D_{\gamma}(E_{\gamma_0}',z)}{D_\theta(z)}\delta = \frac{\delta}{\tau_\theta(E_{\gamma_0},z)}
\end{equation}
where $D_\theta=a(t_E)d_E$ is the angular diameter distance, $\tau_\theta=D_\theta/D_\gamma$ and we have assumed $\tau_\theta> 1, \delta\ll 1$. For small $z$, $\tau_\theta$ is close to  $\tau(E_{\gamma_0},z)$.
Numerically, the above estimate of the size of extended emission around an extragalactic point source is
\begin{equation}
\Theta_{\rm ext}\simeq
\left\{
\begin{array}{ll}
0.5^\circ(1+z)^{-2} \left[\frac{\tau_\theta}{10}\right]^{-1} 
&\\
\left[\frac{\displaystyle E_\gamma}{\displaystyle 0.1\mbox{ TeV}}\right]^{-1}\left[\displaystyle \frac{B_0}{\displaystyle 10^{-14}\mbox{ G}}\right],& \lambda_B'\gg D_e\\
&\\
0.07^\circ (1+z)^{-1/2} \left[\frac{\tau_\theta}{10}\right]^{-1}& \\
\left[\frac{\displaystyle E_\gamma}{\displaystyle 0.1\mbox{ TeV}}\right]^{-3/4}
\left[\frac{\displaystyle B_0}{\displaystyle 10^{-14}\mbox{ G}}\right]
\left[\frac{\displaystyle \lambda_{B0}}{\displaystyle 1\mbox{ kpc}}\right]^{1/2},&\lambda_B'\ll D_e
\end{array}
\right.
\label{Thetaext}
\end{equation}
A clear observational signature of the presence of EGMF induced cascade emission around an initially point source is the decrease of the extension of the source with the increase of photon energy.

Difference in the path between the direct and cascade \gr\ leads to the appearance of delayed emission from a flaring extragalactic \gr\ source. The geometrical scheme of Fig. \ref{fig:scheme} enables to estimate the time delay between the direct photons and secondary inverse Compton \gr s produced by deflected electrons as 
\begin{equation}
T_{\rm delay}=t_{\gamma\gamma}-(t_E-t_{\rm cascade})
\end{equation}
where $t_{\gamma\gamma}$ is found from Eq. (\ref{t_gammagamma}), $t_E$ is the light travel time of direct \gr s from the source,
\begin{equation}
t_E=\int_0^{z_E}\frac{dt}{dz}dz
\end{equation}
and $t_{\rm cascade}$ is the travel time of the secondary IC photons, which is implicitly expressed through the equation
\begin{equation}
\int_{t_0-t_{\rm cascade}}^{t_0} \frac{dt'}{a(t')}=\left(d_E^2+d_{\gamma\gamma}^2-2d_Ed_{\gamma\gamma}\cos(\delta+\Theta_{\rm ext})\right)^{1/2}
\end{equation}
where $t_0$ is the present time. 

In the case $z_{\gamma\gamma}\simeq z_E=z, \delta\ll 1$ the resulting expression is
\begin{equation}
T_{\rm delay}\simeq (1+z)\frac{D_{\gamma}(E_{\gamma_0}',z)\delta^2}{2}\left(1-\frac{D_{\gamma}(E_{\gamma_0}',z)}{D_\theta(z)}\right)
\label{tdelay}
\end{equation}
Numerically, at $z\ll 1$, the time delays for the cases of magnetic fields with large and small correlation lengths are given by
\begin{equation}
T_{\rm delay}\simeq
\left\{
\begin{array}{ll}
7\times 10^5\kappa(1-\tau_\theta^{-1})(1+z)^{-5}&\\
\left[\frac{\displaystyle E_\gamma}{\displaystyle 0.1\mbox{ TeV}}\right]^{-5/2}
\left[\displaystyle \frac{B_0}{\displaystyle 10^{-18}\mbox{ G}}\right]^2\mbox{ s, }& \lambda_B'\gg D_e\\ \\
10^4\kappa(1-\tau_\theta^{-1})(1+z)^{-2}&\\
\left[\frac{\displaystyle E_\gamma}{\displaystyle 0.1\mbox{ TeV}}\right]^{-2}
\left[\frac{\displaystyle B_0}{\displaystyle 10^{-18}\mbox{ G}}\right]^2
\left[\frac{\displaystyle \lambda_{B0}}{\displaystyle 1\mbox{ kpc}}\right]\mbox{ s}, & \lambda_B'\ll D_e
\end{array}
\right.
\label{Tdelay}
\end{equation}
The magnetic field induced time delays could be identified in the observational data via their 
characteristic dependence on the photon energy. 

Magnetic field induced extended emission and delayed "echo" of \gr\ flares are detectable only if the deflection angle $\delta$ is larger than the intrinsic angular scatter of the cascade particles, which is estimated as $\delta_{\rm limit}\simeq m_e/E_e$. Comparing $\delta$ from Eqs. (\ref{delta1}) and (\ref{delta2}) with $\delta_{\rm limit}$ one finds that the measurement of magnetic fields via the detection of cascade emission is possible for $B_0$ larger than
\begin{equation}
\label{limit1}
B_0^{\rm limit}=\left\{
\begin{array}{ll}
\simeq  3 \times 10^{-22} ~\mbox{G}&\\
(1+z)^2\left[\frac{\displaystyle E_\gamma}{\displaystyle 0.1 \mbox{ GeV}}\right]^{1/2}\mbox{ ,}& \lambda_B'\gg D_e\\ &\\
\simeq 1.8\times 10^{-20} ~\mbox{G}~\sqrt{1+z} & \\
\left[\frac{\displaystyle E_\gamma}{\displaystyle 0.1 \mbox{ GeV}}\right]^{1/4}\left[\frac{\displaystyle \lambda_B'}{\displaystyle 1\mbox{ kpc}}\right]^{-1/2}\mbox{ ,} &\lambda_B'\ll D_e.
\end{array}
\right.
\label{blimit}
\end{equation}
From the above equation one can see that lowering the energy threshold of a \gr\ telescope enables to measure weaker magnetic fields. This means that the weakest magnetic fields at the level of the lower bound for the galactic dynamos, $B\sim 10^{-21}$~eV could be best explored with {\it Fermi} which detects photons with energies above $0.1$~GeV.

\section{Lower limit on EGMF strength vs. "universal GeV tails of TeV \gr\ flares" effect}
\label{sec:bound}

\begin{figure*}
\includegraphics[width=\linewidth]{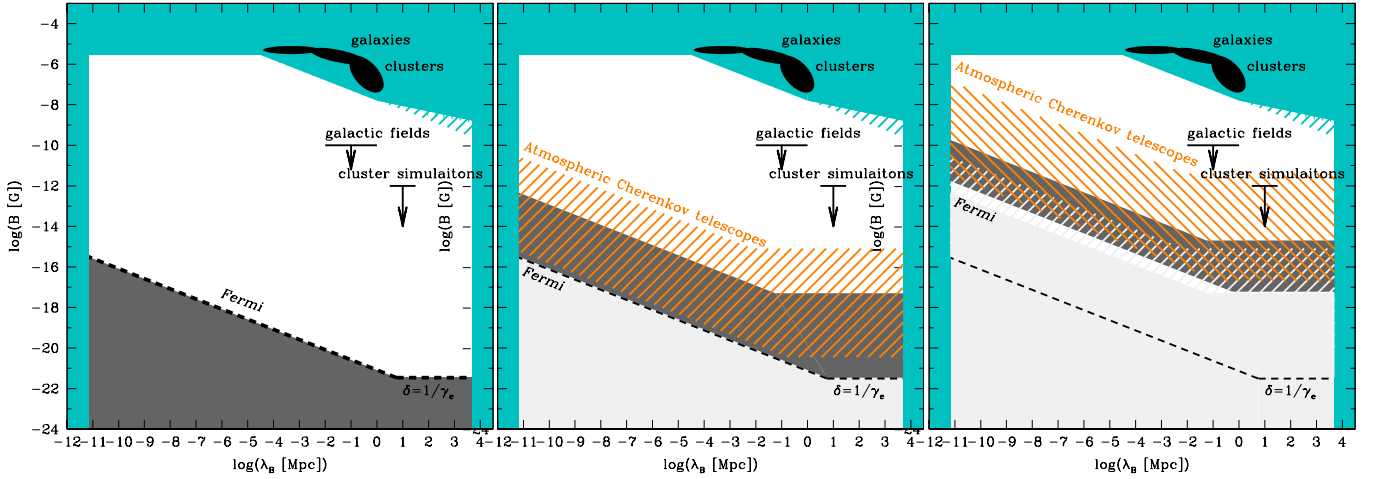}
\caption{Left panel: grey shaded region shows the range of $(B,\lambda_B)$ parameter space which can be excluded via non-observation of "minimal possible" time delay of \gr\ flares by {\it Fermi}. Cyan colored limits on $(B,\lambda_B)$ are the limits shown in details in Figs. \ref{fig:exclusion_experiment} and \ref{fig:exclusion_theory}.  Middle panel:  region of $(B,\lambda_B)$ parameter space which can be probed via observations of time delays of \gr\ flares with {\it Fermi} (dark-grey shaded) and current and next generation atmospheric Cherenkov telescopes (orange hatched). Right panel: region of $(B,\lambda_B)$ parameter space which can be probed via observations of extended emission with {\it Fermi} (dark-grey shaded) and current and next generation atmospheric Cherenkov telescopes (orange hatched). }
\label{fig:exclusion_gamma1}
\end{figure*}

From the discussion of Section \ref{sec:theory} it is clear that  a possibility that the EGMF strength is $B\ll B_{\rm limit}$, where $B_{\rm limit}$ is given by Eq. (\ref{blimit}), is not ruled out observationally and/or theoretically. Moreover, most of the "astrophysical origin" models of the seed fields for the dynamo / compression / turbulence  amplification invoke mechanisms which produce seed fields locally in the parent (proto)galaxy, rather than throughout the IGM, so that the field in the IGM is expected to be nearly zero.

The possibility of the absence of magnetic fields in excess of $B_{\rm limit}$ (\ref{blimit}) in the IGM could be readily tested via observations with {\it Fermi}. Indeed, if the EGMF in the voids of large scale structure does not significantly deflect $e^+e^-$ pairs produced via absorption of the highest energy \gr s from TeV blazars, one still expects that the IC emission from the secondary $e^+e^-$ pairs will produce an observable contribution to the detected \gr\ signal. Moreover, the cascade emission is expected to produce a delayed "tail" of the \gr\ flares, because of the deflections of $e^+e^-$ pairs by an angle $\delta_{\rm limit}\sim m_e/E_e$. Substituting $\delta_{\rm limit}$ at the place of $\delta$ into Eq. (\ref{tdelay}) one finds 
\begin{equation}
\label{Tlimit}
T_{\rm delay,limit}\simeq 10^6 \frac{\kappa(1-\tau_\theta^{-1})}{(1+z)}\left[\frac{E_\gamma}{0.1\mbox{ GeV}}\right]^{-3/2}\mbox{s}
\end{equation}

The time delay of the cascade induced tails of \gr\ flares depends only on the energy of the detected photons and on the distance to the source. This means in the case $B\ll B_{\rm limit}$ one expects to 
\begin{itemize}
\item (a) detect echoes in the $E_\gamma\sim 0.1-1$~GeV energy band  in the flares of all blazars with spectra extending up to the $E_{\gamma_0}\sim 1$~TeV energy band,
\item (b) detect the increase of the time delay of the echoes with the increase of the source redshift and
\item (c) detect the decrease of the time delay of the echoes with the increase of the \gr\ energy.
\end{itemize}

Systematic observation of such delayed "echoes" in {\it Fermi} energy band would be an unambiguous evidence for the hypothesis of extremely weak or zero magnetic fields $B\ll B_{\rm limit}$ in the IGM. At the same time, non-detection of the \gr\ flare echoes with "universal" parameters for all blazars would rule out the possibility of extremely weak magnetic fields or, in other words, impose a lower bound on the EGMF strength, 
\begin{equation}
B\ge B_{\rm limit}.
\end{equation}
and, in this way, rule out a range of "astrophysical origin" or "cosmological origin" models of the seed field.

The curve $B=B_{\rm limit}$ for $E_{\gamma}=0.1$~GeV (typical energy of \gr s observed by {\it Fermi}) is shown by the thick black dashed line in the left panel of Fig. \ref{fig:exclusion_gamma1}.

\section{Measurements of EGMF via time delays in $\gamma$-ray flares}

\subsection{{\it Fermi}}

If $B\ge B_{\rm limit}$, the time delay of the \gr\ flare signal increases above the minimal possible value given by Eq. (\ref{Tlimit}) and starts to depend on $B$. Measuring the (energy dependent) time delay one can, therefore, determine the value of EGMF in the region of the size $R\sim D_\gamma(E_{\gamma_0})$ around the \gr\ source. 

Typical durations of the flares of blazars monitored by {\it Fermi} are $T_{\rm flare}\sim 1-10$~days\footnote{See the {\it Fermi} lightcurves of selected blazars at {\tt http://fermi.gsfc.nasa.gov/ssc/data/access/lat/msl\_lc/}.}. If $\tau(E_{\gamma_0},z)>1$, most of the primary source flux at the energy $E_{\gamma_0}$ is absorbed in interactions with EBL and is subsequently re-emitted within the time $T_{\rm delay}$ at the energy $E_\gamma$. The flux of the delayed emission could be estimated as
\begin{equation}
F_{\rm delay}\sim \frac{T_{\rm flare}}{T_{\rm delay}+T_{\rm flare}}F_0((1+z)E_{\gamma_0}).
\end{equation}

The spectra of all the TeV \gr\ loud blazars have photon indexes harder than $\Gamma= 2$, which corresponds to the equal power emitted per decade of energy. This means that for the TeV blazars 
the energy flux $F_0((1+z)E_{\gamma_0})$ is normally higher than $F_0(E_\gamma)$. This, in turn, implies that for sufficiently distant sources with $\tau(E_{\gamma_0},z)>1$ (significant fraction of the primary source power at the energy $E_{\gamma_0}$ is transferred to the secondary pairs) the delayed cascade flux is comparable or larger than the primary flare flux at the energy $E_\gamma$, as long as $T_{\rm delay}\le T_{\rm flare}$. Thus, if the primary source flare is detectable with {\it Fermi}, the secondary cascade emission should be also readily detectable, at least for the range of time delays $T_{\rm delay}\sim T_{\rm flare}\sim 1-10$~d.

To deconvolve the cascade contribution from the direct emission from the source one has to simultaneously fit the lightcurves in different energy bands with a sum of direct time dependent emission from the source $\exp\left[-\tau(E_\gamma,z)\right]F_0(E_\gamma,t)$ plus a delayed contribution. The expected lightcurve of the delayed emission is obtained via a convolution of the direct emission lightcurve with an exponential kernel \cite{japanese}
\begin{eqnarray}
\label{flux}
F_{\rm delay}(t,E_\gamma)=\int_{-\infty}^t \left(1-e^{-\tau(E_{\gamma_0},z)}\right)F_0((1+z)E_{\gamma_0},t')\nonumber\\
\exp\left[-\frac{t-t'}{T_{\rm delay}(E_\gamma)}\right]dt'
\end{eqnarray}

The GeV \gr s are emitted by electrons/positrons produced by the primary \gr s of the energies $E_{\gamma_0}\simeq 1.3$~TeV (see Eq. (\ref{Esec})). Such \gr s can travel over several hundred Mpc distances (see Eq. (\ref{Dgamma0})). This means that the timing observations with {\it Fermi} would reveal the magnetic fields averaged over many voids of the large scale structure.

To estimate the maximal magnetic field detectable via the measurements of time delays, we notice that if the time delay becomes much larger than the duration of the flare, the delayed flux diminishes by a factor  $T_{\rm delay}/T_{\rm flare}\gg 1$, compared to the primary source flux (see Eq. (\ref{flux})). If the delayed flux becomes much smaller than the quiescent source flux, the delayed emission is difficult to detect. Assuming that the flux enhancement in the 1-10~d time scale flares of {\it Fermi} blazars is by a factor of $\sim 10$, we can estimate that the maximal detectable time delays are of the order of $T_{\rm delay}\sim 10^2$~days. Since the EGMF induced time delay is smallest at the largest energies, the maximal time delay is best detectable at the highest energies accessible for the observations, which are $E_\gamma\sim 100$~GeV for {\it Fermi}. Substituting the estimated maximal $T_{\rm delay}$ and $E_{\gamma}$ into Eq. (\ref{Tdelay}) one finds that the maximal magnetic field is
\begin{equation}
\label{bmax_TFermi}
B_{0}^{\rm max}\simeq \left\{
\begin{array}{ll}
4\times 10^{-18}\kappa^{-1/2}(1-\tau_\theta^{-1})^{-1/2}(1+z)^{5/2}& \\ \left[\frac{\displaystyle E_\gamma}{\displaystyle 0.1\mbox{ TeV}}\right]^{5/4}\left[\frac{\displaystyle T_{\rm delay}}{\displaystyle 10^2\mbox{ d}}\right]^{1/2}\mbox{ G}, & \lambda_B\gg D_e\\
&\\
3\times 10^{-17}\kappa^{-1/2}(1-\tau_\theta^{-1})^{-1/2}(1+z)&\\
\left[\frac{\displaystyle E_\gamma}{\displaystyle 0.1\mbox{ TeV}}\right]\left[\frac{\displaystyle \lambda_{B,0}}{\displaystyle 1\mbox{ kpc}}\right]^{-1/2}\left[\frac{\displaystyle T_{\rm delay}}{\displaystyle 10^2\mbox{ d}}\right]^{1/2}\mbox{ G}, & \lambda_B\ll D_e
\end{array}
\right.
\end{equation}
The range of EGMF strengths measurable via time delays with {\it Fermi} is shown as the lower dark grey shaded region in the middle panel of Fig. \ref{fig:exclusion_gamma1}.

\subsection{Cherenkov telescopes}

Ground based Cherenkov telescopes detect \gr s of somewhat higher energies ($E_\gamma\ge 50$~GeV) than {\it Fermi}.  Next generation Cherenkov telescopes, like CTA, are expected to reach the low energy threshold of about $E_\gamma\sim 10$~GeV. Higher threshold energy of Cherenkov telescopes leads to somewhat higher limiting magnetic field (\ref{blimit})
\begin{equation}
\label{bmin_TCTA}
B_0^{\rm limit}\simeq \left\{
\begin{array}{ll}
3\times 10^{-21}(1+z)^2\left[\frac{\displaystyle E_\gamma}{\displaystyle 10\mbox{ GeV}}\right]& \lambda_B\gg D_e\\
&\\
6\times 10^{-20}\sqrt{1+z}\left[\frac{\displaystyle E_\gamma}{\displaystyle 10\mbox{ GeV}}\right]^{1/4}&\\
\left[\frac{\displaystyle \lambda_B(z)}{\displaystyle 1\mbox{ kpc}}\right]^{-1/2}\mbox{ G}, & \lambda_B\ll D_e
\end{array}
\right.
\end{equation}
and much shorter minimal delay time, $T_{\rm limit}\simeq 10^3\kappa(1-\tau_\theta^{-1})(1+z)^{-1}\left[E_\gamma/10 \mbox{ GeV}\right]^{-3/2}$~s. 

It is clear that in the absence of magnetic fields in excess of $B_{\rm limit}$ one expects to observe the "universal GeV tails of TeV flares" effect, discussed in the previous section also with the next generation Cherenkov telescopes. The prospects for detection and study of such "universal tails" with Cherenkov telescopes are potentially even better than with {\it Fermi}. The effective collection area of the ground based \gr\ telescopes at the low energy threshold is at least 4 orders of magnitude larger than the effective area of {\it Fermi}. This enables detection of very short time delays. In fact, the measurements of time delay between the signal of blazar \gr\ flares in different energy bands reach $\sim 10$~s already with the current generation instruments \cite{MAGIC_QG,HESS_QG}. At the energies above $\sim 10$~GeV the minimal measurable time delays with the next-generation instruments will be limited mostly by the intrinsic time spread of the \gr\ flare, rather then by the sensitivity of the instrument. At the moment, fastest observed variability time scale of blazar flares is $\sim 1$~min, comparable to the minimal possible variability time scale, given by the light-crossing time of the blazar's central engine, the supermassive black hole \cite{HESS_PKS,MAGIC_501,neronov08}. Assuming that the minimal measurable time delays are $T_{\rm delay}\sim 10$~s, one can check that the fields of the strength at the level given by Eq. \ref{bmin3}) could be readily measured. 

Re-scaling the estimate of the maximal measurable magnetic field (\ref{bmax_TFermi}) to the higher energies accessible for Cherenkov telescopes one finds that the timing measurements are sensitive to the magnetic fields with the strength up to
\begin{equation}
\label{bmax_TCTA}
B_{0}^{\rm max}\simeq \left\{
\begin{array}{ll}
10^{-15}\kappa^{-1/2}(1-\tau_\theta^{-1})^{-1/2}(1+z)^{5/2}& \\ \left[\frac{\displaystyle E_\gamma}{\displaystyle 10\mbox{ TeV}}\right]^{5/4}\left[\frac{\displaystyle T_{\rm delay}}{\displaystyle 10^2\mbox{ d}}\right]^{1/2}\mbox{ G}, & \lambda_B\gg D_e\\
&\\
3\times 10^{-15}\kappa^{-1/2}(1-\tau_\theta^{-1})^{-1/2} (1+z) &\\
\left[\frac{\displaystyle E_\gamma}{\displaystyle 10\mbox{ TeV}}\right]
\left[\frac{\displaystyle \lambda_{B,0}}{\displaystyle 1\mbox{ kpc}}\right]^{-1/2}\left[\frac{\displaystyle T_{\rm delay}}{\displaystyle 10^2\mbox{ d}}\right]^{1/2}\mbox{ G}, & \lambda_B\ll D_e
\end{array}
\right.
\end{equation}

The range of EGMF strengths measurable via time delays with Cherenkov telescopes is shown as the orange hatched region in the middle panel of Fig. \ref{fig:exclusion_gamma1}.

\section{Measurement of EGMF via detection of extended emission around extragalactic point sources}

\subsection{{\it Fermi}}

If the magnetic field is large enough, deflections of the cascade $e^+e^-$ pairs can be detected not only via time delay of the cascade emission, but also as extended emission of the angular size  (\ref{Thetaext}) directly in the images of the extragalactic \gr\ sources. 

The extension of the source could be detected if the source size is larger than the point spread function (PSF) of a telescope. The minimal magnetic field measurable via the study of extended emission is the one at which the extension of the source becomes larger than the telescope's PSF at low energy threshold. 

In the particular example of {\it Fermi}, the PSF depends on the photon energy, 
decreasing roughly as $\Theta_{\rm PSF}\simeq 0.3^\circ\left[E_\gamma/1\mbox{ GeV}\right]^{-0.8}$ below $E_\gamma\simeq 1$~GeV and slowly improving from $\sim 0.3^\circ$ at $1$~GeV to $\Theta_{\rm PSF}\simeq 0.1^\circ$ at $E_\gamma\sim 100$~GeV.

Substituting the reference energy $E_\gamma=1$~GeV and $\Theta_{\rm ext}=0.3^\circ$ into Eq. (\ref{Thetaext}) one can find the minimal field measurable with {\it Fermi}
\begin{equation}
\label{bmin3}
B_{0}^{\rm min}\simeq \left\{
\begin{array}{ll}
6\times 10^{-17}\frac{\tau_\theta}{10} (1+z)^2\left[\frac{\displaystyle E_\gamma}{\displaystyle 1\mbox{ GeV}}\right]&\\ \left[\frac{\displaystyle \Theta_{\rm ext}}{\displaystyle 0.3^\circ}\right]\mbox{ G}, & \lambda_B\gg D_e\\
1.3\times 10^{-15}\tau_\theta (1+z)^{1/2}\left[\frac{\displaystyle E_\gamma}{\displaystyle 1\mbox{ GeV}}\right]^{3/4}&\\
\left[\frac{\displaystyle \lambda_{B0}}{\displaystyle 1\mbox{ kpc}}\right]^{-1/2}\left[\frac{\displaystyle \Theta_{\rm ext}}{\displaystyle 0.3^\circ}\right]\mbox{ G}, & \lambda_B\ll D_e
\end{array}
\right.
\end{equation}
It is interesting to note that, in principle, the size of the extended source grows with the decrease of energy below $E_\gamma\sim 1$~GeV, so that, apparently, it should be easier to detect the extension of the source at lower energies, close to the low-energy threshold of {\it Fermi}, $E_\gamma\sim 0.1$~GeV. However, for {\it Fermi}, $\Theta_{\rm  PS}$ grows with the decrease of energy below $E_\gamma\sim 1$~GeV roughly in the same way as $\Theta_{\rm ext}$, so that the decrease of the photon energy does not facilitates the detection of extended emission.

The total flux of the extended source does not depend on the magnetic field strength. This means that the surface brightness of the source decreases inversely proportional to the $\Theta_{\rm ext}^2$. Detection of extended emission becomes impossible when the surface brightness of the extended source becomes comparable to the fluctuations of the diffuse background. If the minimal point source flux detectable by telescope is $F_{\rm PSF}$, the minimal detectable extended source flux scales with the source size as 
$F_{\rm ext}\simeq \left(\Theta_{\rm ext}/\Theta_{\rm PSF}\right)^2$. For {\it Fermi}, $F_{\rm PS}$ depends on the photon energy approximately as $F_{\rm PS}\simeq 10^{-12}\left(E_\gamma/1\mbox{ GeV}\right)$~erg/cm$^2$s at $E_\gamma\gg 1$~GeV.

The flux of the brightest TeV blazars reaches $F_0\sim 10^{-10}$~erg/cm$^2$s.   If the source is at sufficiently large distance, so that $\tau_\theta(E_{\gamma_0},z)\ge 1$, the extended source flux at the energy $E_\gamma$ is $F_{\rm ext}\sim F_0$. Such flux can be detected by {\it Fermi} up to the energies $E_\gamma\sim 100$~GeV is the source size is not larger than $\Theta_{\rm ext}\le 0.1^\circ\left(E_\gamma/100\mbox{ GeV}\right)^{-1}$. This enables to estimate the maximal magnetic field  measurable with {\it Fermi} as
\begin{equation}
\label{bmin4}
B_{0,\rm max}\simeq \left\{
\begin{array}{ll}
2\times 10^{-15}(1+z)^2&\\ \left[\frac{\displaystyle \tau_\theta}{\displaystyle 10}\right]\left[\frac{\displaystyle \Theta_{\rm ext}}{\displaystyle 0.1^\circ}\right]\mbox{ G}, & \lambda_B\gg D_e\\
1.4\times 10^{-14}(1+z)^{1/2}\left[\frac{\displaystyle E_\gamma}{\displaystyle 100\mbox{ GeV}}\right]^{3/4}&\\
\left[\frac{\displaystyle \lambda_{B0}}{\displaystyle 1\mbox{ kpc}}\right]^{-1/2}\left[\frac{\displaystyle \Theta_{\rm ext}}{\displaystyle 0.1^\circ}\right]\left[\frac{\displaystyle \tau_\theta}{\displaystyle 10}\right]\mbox{ G}, & \lambda_B\ll D_e
\end{array}
\right.
\end{equation}

The range of magnetic fields measurable via a study of extended emission around extragalactic point source with {\it Fermi} is shown as an upper dark shaded region in the right panel of  Fig. \ref{fig:exclusion_gamma1}.

\subsection{Cherenkov telescopes}

Angular resolution of Cherenkov telescopes is comparable to that of {\it Fermi} at the energies $E_\gamma\sim 100$~GeV. Somewhat higher low energy threshold of Cherenkov telescopes results in a higher estimate of the minimal magnetic field strength measurable via detection of extended emission,
\begin{equation}
\label{bmin5}
B_{0}^{\rm min}\simeq \left\{
\begin{array}{ll}
2\times 10^{-16}\tau_\theta(1+z)^2 \left[\frac{\displaystyle E_\gamma}{\displaystyle 10 \mbox{ GeV}}\right]&\\ \left[\frac{\displaystyle \Theta_{\rm ext}}{\displaystyle 0.1^\circ}\right]\mbox{ G}, & \lambda_B\gg D_e\\
2.5\times 10^{-15}\tau_\theta (1+z)^{1/2} \left[\frac{\displaystyle E_\gamma}{\displaystyle 10\mbox{ GeV}}\right]^{-1/4}&\\
\left[\frac{\displaystyle \lambda_B}{\displaystyle 1\mbox{ kpc}}\right]^{-1/2}\left[\frac{\displaystyle \Theta_{\rm ext}}{\displaystyle 0.1^\circ}\right]\mbox{ G}, & \lambda_B\ll D_e
\end{array}
\right.
\end{equation}

Much larger effective collection area of Cherenkov telescopes explains their better performance at the highest energies. Assuming that photons of the energy up to $\sim 6$~TeV (produced by the primary \gr s of the energies $E_{\gamma_0}\simeq 100$~TeV could be detected, one can estimate the maximal EGMF strength from the condition that the size of the extended source should be smaller than the size of the telescope's field of view (FoV)
\begin{equation}
\label{bmin6}
B_{0}^{\rm max}\simeq \left\{
\begin{array}{ll}
3.6\times 10^{-12}(1+z)^2\left[\frac{\displaystyle E_\gamma}{\displaystyle 6\mbox{ TeV}}\right]\left[\frac{\displaystyle \tau_\theta}{\displaystyle 10}\right]&\\ \left[\frac{\displaystyle \Theta_{\rm ext}}{\displaystyle 3^\circ}\right]\mbox{ G}, & \lambda_B\gg D_e\\
9\times 10^{-12}(1+z)^{1/2}\left[\frac{\displaystyle E_\gamma}{\displaystyle 6\mbox{ TeV}}\right]^{3/4}&\\
\left[\frac{\displaystyle \lambda_{B0}}{\displaystyle 1\mbox{ kpc}}\right]^{-1/2}\left[\frac{\displaystyle \Theta_{\rm ext}}{\displaystyle 3^\circ}\right]\left[\frac{\displaystyle \tau_\theta}{\displaystyle 10}\right]\mbox{ G}, & \lambda_B\ll D_e
\end{array}
\right.
\end{equation}
Typical sizes of the FoV of Cherenkov telescopes are several degrees. Obviously, extension of the FoV up to $\sim 10^\circ$, expected e.g. with the high-energy part of CTA or with AGIS will extend the range of EGMF strengths accessible for the study toward stronger fields. 

The range of the field strengths measurable via study of extended emission around extragalactic sources with Cherenkov telescope is shown at the red right-hatched region in the right panel of Fig. \ref{fig:exclusion_gamma1}.

\section{Evidence for strong EGMF}
\label{sec:strongB}

If the search of EGMF with $(B,\lambda_B)$ in the range outlined by the grey shaded region in Fig. \ref{fig:exclusion_theory} will not give positive results and the lower bound on EGMF strength discussed 
in Section \ref{sec:bound} would be confirmed, this would signify that magnetic fields in excess of $\sim 10^{-12}$~G are present all over the IGM. A direct test of this result will be possible with the help of imaging of extragalactic \gr\ sources. Strong EGMF should cause a phenomenon of the presence of "universal" extended pair halos around {\it all} extragalactic multi-TeV \gr\ sources \cite{coppi}, with properties which depend on the source redshift, but do not depend on the EGMF strength. 

Namely, if the magnetic field is strong enough, trajectories of $e^+e^-$ pairs are strongly deflected over the cooling length $D_e$, so that the deflection angle $\delta$ becomes $\delta\sim 1$. In this case the secondary IC photons are emitted isotropically from a region of the size $R\sim D_\gamma(E_{\gamma_0})$ around the primary \gr\ source. If the primary source emits \gr s isotropically, the expected flux of the  extended halo at the energy $E_\gamma$ is equal to the primary unabsorbed source flux at the energy $E_{\gamma_0}$. The angular size of the halo, $\theta_{\rm halo}(E_\gamma)\sim D_\gamma(E_{\gamma_0})/D\simeq 1/\tau_\theta(E_{\gamma_0},z)$, depends on the source redshift, but not on the intrinsic properties of the source or on the EGMF strength. 

If the primary source (e.g. a blazar) emits the primary \gr\ flux anisotropically, in a jet with an opening angle $\Theta_{\rm jet}$, the cascade emission fills a cone with an opening angle $\Theta_{\rm jet}$ and height $H\sim D_\gamma(E_{\gamma_0})$, rather then forms a spherically symmetric halo around the source. In this case the size of extended source produced by the cascade emission is expected to be smaller, $\theta_{\rm halo}\sim D_\gamma(E_{\gamma_0})\Theta_{\rm jet}/\left(D-D_\gamma(E_{\gamma_0})\right)\simeq \Theta_{\rm jet}/(\tau_\theta(E_{\gamma_0},z)-1)$. At the same time, the flux in the extended source is expected to be suppressed by a factor $\Theta_{\rm jet}^2$, compared to the primary source flux at the energy $E_{\gamma_0}$. 
 
The halos around sources which are distant enough, $\tau(E_{\gamma_0},z)\gg 1$, and having significant fluxes at high energies, $F_0(E_{\gamma_0})\sim F(E_\gamma)$, should be readily detectable with the current and next generation VHE \gr\ telescopes. In fact, if the primary source flux is characterized by the photon index $\Gamma<2$ the extended halo flux could dominate over the primary source flux at the energy $E_\gamma$, in the case of an isotropic primary source and be moderately below the point source flux in the case of an anisotropic primary source. For example, the flux from the brightest TeV blazars is at the level $F_0(E_{\gamma_0})\sim 10^{-10}$~erg/cm$^2$s at the energies $E_{\gamma_0}\sim 1-10$~TeV. If the primary emission from the source is beamed into a jet with an opening angle $\Theta_{\rm jet}\sim 5^\circ$ (plausible assumption which implies the bulk Lorentz factor $\sim 10$ for the blazar jet), the expected EGMF-independent extended emission flux at the energies 
i.e. at the level detectable by {\it Fermi}.

The part of the $(B,\lambda_B)$ parameter space in which the existence of non-zero EGMF could be established via imaging \gr\ observations, but no measurement of EGMF is possible, is shown as an upper light-grey shaded region in Fig. \ref{fig:exclusion_theory}.

\section{Conclusions}
\label{sec:conclusions}

We have discussed the prospects of detection of weak magnetic fields in the intergalactic medium with the novel techniques of timing and imaging observations with ground and space-based \gr\ telescopes.
These techniques enable measurements of extremely weak magnetic fields with strengths much lower than the ones accessible for the measurements with radio telescopes (via Zeeman splitting and/or Faraday rotation techniques). 

We have demonstrated that using \gr\ observations one can detect, or rule out the possibility of existence of cosmologically or astrophysically produced "seed" magnetic fields in the voids of the large scale structure, which are conjectured to exist in a range of theories of the origin of magnetic fields in galaxies and galaxy clusters (see Fig. \ref{fig:exclusion_theory}). 

To summarize, we find that discovery or non-detection of weak magnetic fields  in the voids of the large scale structure should provide, in the nearest future, a decisive test of the theories of the origin of cosmic magnetic fields.

\section*{Acknowledgement}

We would like to thank A.~Dolgov, R.~Durrer, K.~Jedamzik, P.~Kronberg and T.Vachaspati for fruitful discussions of the subject and comments on the text of the Manuscript.
AN would like to thank LUTH department of Paris Observatory for hospitality during visit in the course of which the work on the present paper started. The work of AN is supported by the Swiss National Research Foundation (grant PP00P2\_123426).

\end{document}